\documentclass[aps,prl,twocolumn,floatfix,superscriptaddress]{revtex4}

\usepackage{epsfig}
\usepackage{amsmath}

\begin{document}
\preprint{DUKE-TH-02-230}

\title{Semi-hard scattering of partons at SPS and RHIC: A study in 
contrast}
\author{Steffen A.~Bass}
\affiliation{Department of Physics, Duke University, 
             Durham, NC 27708-0305}
\affiliation{RIKEN BNL Research Center, Brookhaven National Laboratory, 
	Upton, NY 11973, USA}
\author{Berndt M\"uller}
\affiliation{Department of Physics, Duke University, 
             Durham, NC 27708-0305} 
\author{Dinesh K.~Srivastava}
\altaffiliation{on leave from: Variable Energy Cyclotron Centre, 
             1/AF Bidhan Nagar, Kolkata 700 064, India}            
\affiliation{Physics Department, McGill University,
             3600 University Street, Montreal, H3A 2T8, Canada} 
\date{\today}
\begin{abstract}
We analyze the contribution of pQCD based 
semi-hard parton (re)scattering to the reaction
dynamics of relativistic heavy-ion collisions at SPS and RHIC.
While such processes are able to account for
the measured yield of high momentum direct photons at SPS energies, 
the conditions necessary for jet-quenching are not fulfilled. 
The situation changes dramatically at RHIC energies.
\end{abstract}
\maketitle
The detection of a  quark gluon plasma would confirm one of the
important predictions of QCD, namely the de-confinement of quarks 
and gluons when nuclear matter is heated to a temperature exceeding 
$T_c \approx 170$ MeV.  Several of the proposed signatures of the 
formation of a quark-gluon plasma \cite{qgp_rev} in relativistic 
heavy ion collisions have been observed in collisions of lead nuclei 
at the CERN SPS.  Even though explanations in terms of solely hadronic 
interactions for many of these observations have been proposed from time
to time, the simplest and most consistent interpretation of the
observations is in terms of the formation of a de-confined state 
\cite{signs}.  Microscopic calculations confirm that the energy 
density reached in these collisions is, indeed, high enough for 
quark de-confinement to occur \cite{weber98}.

The  success~\cite{xnw} of  perturbative QCD in describing the 
transverse momentum spectrum of high-$p_T$ pions measured in Pb+Pb
collisions at the SPS~\cite{pions} suggests that the perturbative 
scattering of nuclear partons contributes to the energy deposition
in these reactions. On the other hand, the apparent absence of a 
visible energy loss by these partons, in stark contrast to the
observations at RHIC~\cite{jet-quench}, indicates a dearth of
multiple scattering that could lead to local equilibration of 
the partons before hadrons are formed.  

Prompt photons are a good probe of these phenomena, since they escape
from the strongly interacting matter without major distortion by final 
state interactions.  We have recently shown that the measured photon 
yield can be related to the number of semi-hard scatterings among the
partons, because a substantial fraction of the photons are emitted in
secondary interactions of scattered partons \cite{fms,bms_phot}.  

In the present study we report the results of a parton cascade model 
calculation 
of photon emission in central collision of Pb nuclei at CERN-SPS 
energies.  We address the following questions:
\begin{itemize}
\item How large is the contribution of pQCD based parton scattering
	to the reaction dynamics of nuclear collisions at the SPS?
\item Can this mechanism account for the direct photon yield as 
	measured by the WA98 experiment~\cite{wa98}?
\item Does the perturbative scattering of partons create conditions 
	necessary for jet quenching, and how do these conditions
	compare to those predicted by the same model for collisions
	at RHIC energies?
\end{itemize}

The parton cascade model (PCM) provides a detailed 
space-time description of nuclear collisions at high energy, 
from the onset of hard interactions among the partons of the 
colliding nuclei up to the moment of hadronization~\cite{GM}. 
It is based on the relativistic Boltzmann equation for the time 
evolution of the parton density due to perturbative QCD interactions:
\begin{equation}
p^\mu \frac{\partial}{\partial x^\mu} F_i(x,p) = {\cal C}_i[F] \, .
\label{eq03}
\end{equation}
The PCM collision term ${\cal C}_i$ is a nonlinear functional of the 
phase-space distribution function $F(x,p)$, containing the matrix 
elements which account for the following processes:
\begin{equation}
\label{processes}
\begin{array}{lll}
g g \to g g \quad&\quad g g \to q \bar q 
  &\quad q g \to q g \\
q q' \to q q' \quad&\quad q q \to q q 
  &\quad q \bar q \to q' \bar q' \\
q \bar q \to q \bar q 
  &\quad q \bar q \to g g \quad& \\
q g \to	q \gamma \quad&\quad q \bar q \to \gamma \gamma 
  & \quad q \bar q \to g \gamma
\end{array}
\end{equation}
with $q$ and $q'$ indicating different quark flavors.  The 
corresponding scattering cross sections are expressed in terms
of spin- and color-averaged amplitudes $|{\cal M}|^2$~\cite{Cutler.78}:
\begin{equation}
\label{dsigmadt}
\left( \frac{{\rm d}\hat \sigma}
     {{\rm d} Q^2}\right)_{ab\to cd} \,=\, \frac{1}{16 \pi \hat s^2}
        \,\langle |{\cal M}|^2 \rangle
\end{equation}
The total cross section, necessary for the transport calculations is
obtained from
(\ref{dsigmadt}):
\begin{equation}
\label{sigmatot}
\hat \sigma_{ab}(\hat s) \,=\, 
\sum\limits_{c,d} \, \int\limits_{(p_T^{\rm min})^2}^{\hat s}
        \left( \frac{{\rm d}\hat \sigma }{{\rm d} Q^2}
        \right)_{ab\to cd} {\rm d} Q^2 \quad .
\end{equation}
The low momentum-transfer cut-off $p_T^{\text{min}}$ regularizes 
the infrared divergence of the parton-parton cross section.  We 
have used a value of  $Q_0^2=(p_T^{\text{min}})^2$.
We choose the GRV-HO parameterization \cite{grv} and sample the
distribution functions at the initialization scale $Q_0^2$.

Additionally, we include the branchings $q \to q g$, $q \to q\gamma$, 
$ g \to gg$ and $g \to q\overline{q}$ \cite{frag}.  The soft and 
collinear singularities in the showers are avoided by terminating the
branchings when 
the virtuality of the time-like partons drops below $\mu_0 = 1$ GeV.
Some of these aspects were discussed in~\cite{SG}
and explored within framework of the PCM implementation VNI~\cite{vni}. 
The present work is based on our thoroughly revised, corrected, and
extensively tested implementation of the parton cascade model, 
called VNI/BMS~\cite{bms_big1}.

\begin{table}[tb]
\renewcommand{\arraystretch}{1.2}
\begin{tabular}{|l|r|r|}
  \hline
   & SPS & RHIC \\\hline\hline
cut-off $p_T^{\rm {min}}$ (GeV) & 0.7 & 1.0 \\
\# of hard collisions/event & 255 & 3618 \\
\# of fragmentations/event  & 17  & 2229 \\
average momentum transfer $\langle Q^2 \rangle$ (GeV$^2$) & 0.8 & 1.7\\
average c.m. energy $\langle \sqrt{\hat{s}} \rangle$ (GeV) & 2.6 & 4.7\\
\hline
\end{tabular}
\caption{Comparison between the total number of 
semi-hard partonic collisions and 
fragmentations in Pb+Pb collisions at SPS and Au+Au collisions at RHIC.}
\end{table}

The starting point of our investigation is the comparison of the number
and average properties of perturbative parton-parton collisions between 
nuclear reactions at the SPS and RHIC, displayed in table~1. 
Following \cite{eskola2k} we choose 
a low-momentum cut-off of 0.7~GeV for our SPS calculation and 1.0~GeV for
the RHIC calculation. The number of perturbative parton-parton scatterings
at SPS is on average 225 per event -- clearly too small to provide a 
significant effect on the dynamics and characteristics of the bulk of the
matter. The number of perturbative
scatterings increases by a factor of 14 from SPS to RHIC -- at RHIC these
interactions are as important for the collision dynamics as the soft,
non-perturbative processes. 
The number of fragmentations increases likewise by more than two orders 
of magnitude -- at SPS it is virtually negligible and has no influence
on the parton collision cascade. 
The final entries of table~1 refer to the average momentum transfer 
$\langle Q^2 \rangle$ in semi-hard collisions and their average center
of mass energy $\sqrt{\hat s}$. For the SPS, 
both values are clearly cut-off dominated.
The lack of fragmentations at the SPS and the small values for
the average c.m. energy and momentum transfer
are a clear indication that the conditions for jet-quenching to occur are 
not fulfilled at this energy.

\begin{figure}[tb]   
  \begin{center}
  \epsfig{file=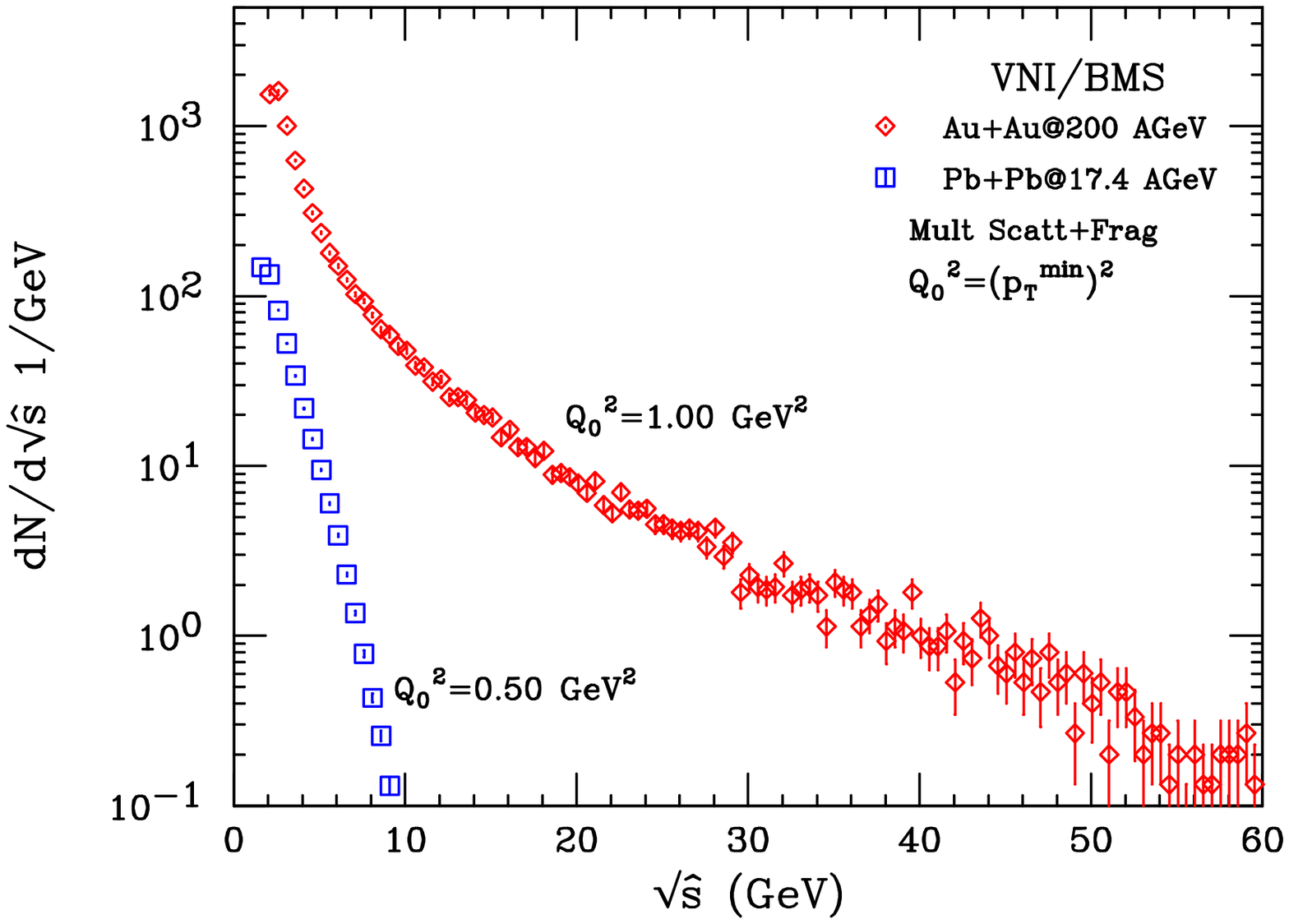,width=8.6cm}
  \epsfig{file=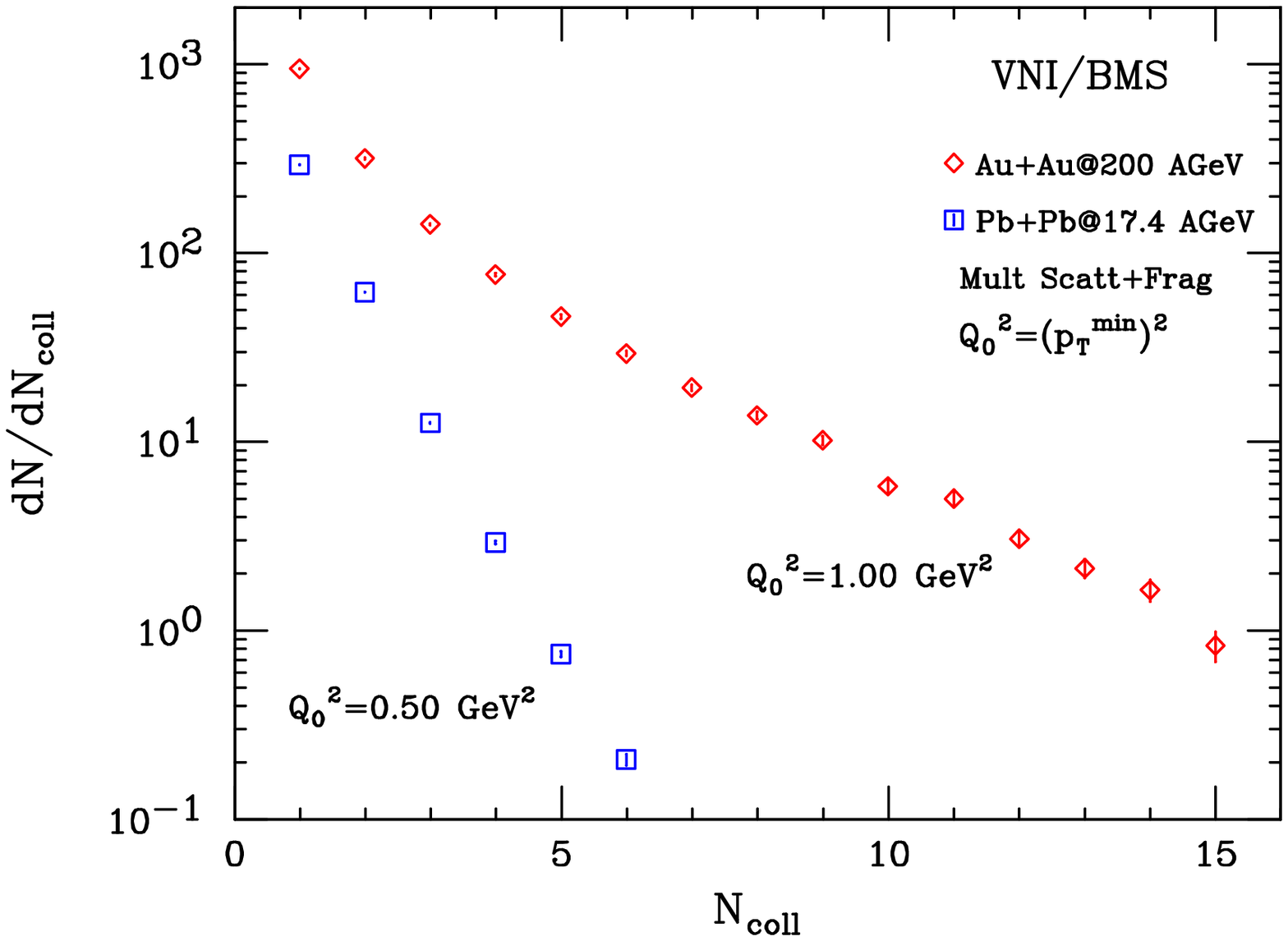,width=8.6cm}
\caption{ 
Upper panel: 
  The distribution of center of mass energies of partonic collisions at 
  RHIC and SPS.
Lower panel: 
  Distribution of the number of collisions suffered by a parton at 
  RHIC and SPS.}
\label{fig1}
\end{center}
\end{figure}

A more detailed  comparison between the regimes at SPS and RHIC 
can be seen in Fig.~\ref{fig1}. The upper frame shows the $\sqrt{\hat{s}}$ 
distribution for binary parton-parton collisions at SPS and RHIC. 
Obviously the collisional regime at SPS is much softer than at 
RHIC reflecting the strong suppression in the number of semi-hard
scatterings observed at the SPS. The two distributions differ strongly
not only in magnitude, but also in shape
-- the distribution at RHIC cannot be obtained by simply rescaling the 
the distribution for the SPS energy and exhibits strong enhancement
towards higher values of $\sqrt{\hat s}$.
The bottom frame of Fig.~\ref{fig1} 
shows a comparison of the collision number distributions for SPS and 
RHIC. Although multiple parton-parton scatterings occur at SPS energy, 
such re-scattering is strongly suppressed compared to RHIC.  E.g., the 
probability for a parton to scatter five times is more than two orders 
of magnitude larger at RHIC than at the SPS, and the difference between
the two curves grows with the number of multiple scatterings.

\begin{figure}[tb]
\begin{center}  
\epsfig{file=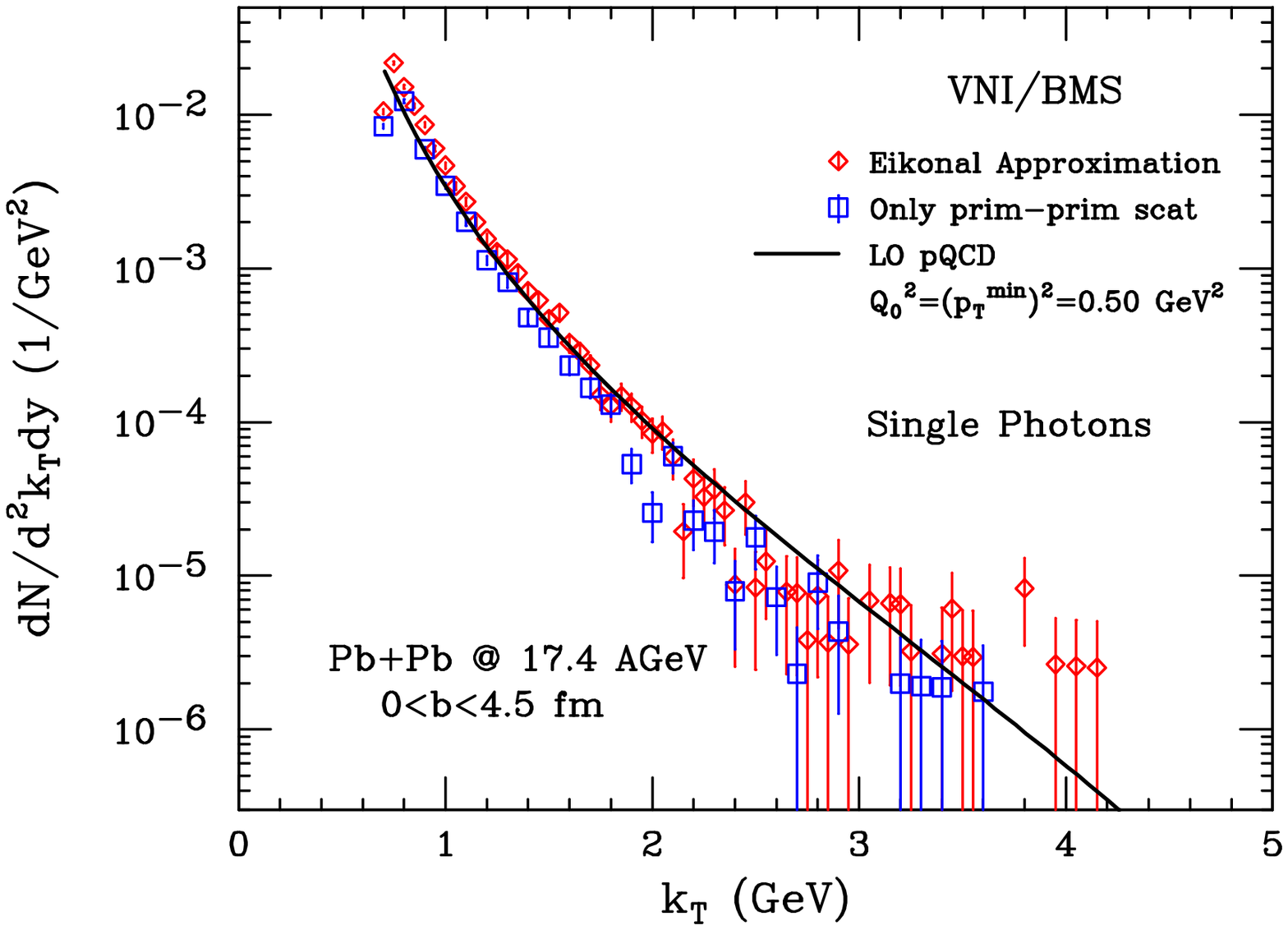,width=8.2cm}
\epsfig{file=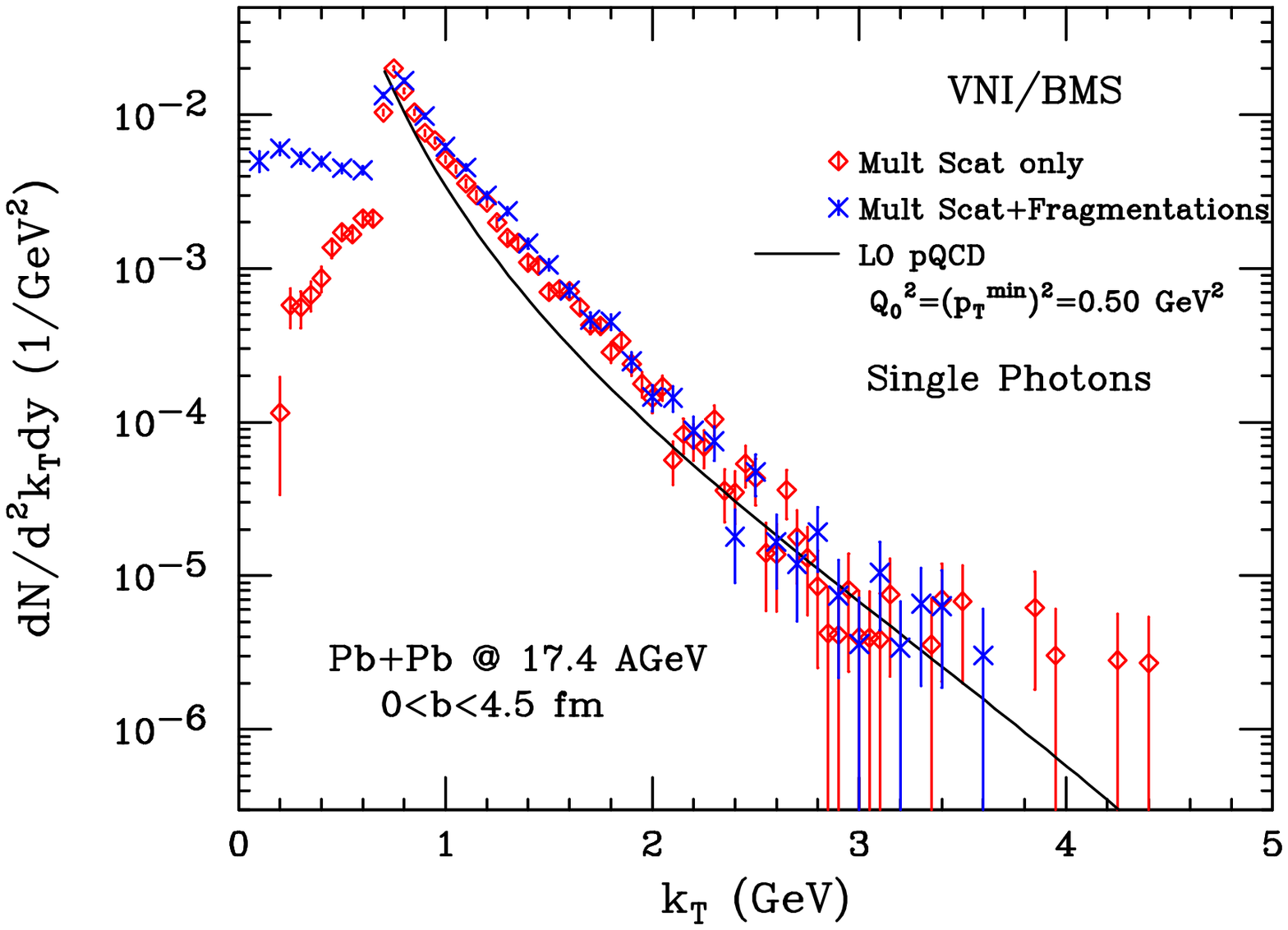,width=8.2cm}
\caption{ 
Upper panel: 
  Transverse momentum distribution of photons from central
  collision of lead nuclei at SPS in lowest order pQCD (solid line)
  in the eikonal approximation (diamonds), and for primary collisions 
  only (squares), generated with VNI/BMS.
Lower panel:
  Photon transverse momentum spectra for multiple scatterings only 
  (diamonds) and for multiple scatterings and fragmentations 
  ($\times$ symbols). The solid line shows the predictions of 
  lowest-order perturbative QCD.}
\label{fig2}
\end{center}
\end{figure}

In the following, we shall focus on photon production at the SPS, 
since high-$k_T$ photons constitute a direct probe of semi-hard
parton scattering \cite{bms_phot}. 
As a first step, we discuss invariant transverse momentum spectra 
of photons obtained when performing the PCM calculation within the 
eikonal approximation, as well as with the restriction to primary 
interactions. These should correspond to a lowest order pQCD estimate 
(see upper frame of Fig.~\ref{fig2}).  We consider near 
central collision of Pb nuclei with impact parameters $0<b<4.5$ fm, 
corresponding to the centrality covered in the WA98 experiment. All
the results are for the region of the central rapidity. 
The close agreement between these three different calculations 
confirms their insensitivity to the assumptions as well as
the accuracy of our numerical implementations. Deviations from 
these results would then provide evidence for multiple scattering. 

The lower frame of Fig.~\ref{fig2} shows calculations when (a) only 
multiple scattering among the partons, and (b) multiple scatterings as 
well as fragmentations of the partons are included.  Multiple 
scattering of partons leads to a broadening of the $k_T$ distribution
for intermediate $k_T$ in the range 1.0~GeV~$\le k_T \le$~2.5~GeV.
We find that at SPS the fragmentations do {\em not} result in an increasing 
number of collisions. The increase in photon production, all due 
to the fragmentation $q \to q\gamma$,  is of the of the order of 30\%.
This result contrasts sharply with the results~ \cite{bms_phot} obtained 
for collisions at RHIC where fragmentation processes enhance multiple 
scatterings by a factor of almost 2.4 (in the absence of LPM suppression).

\begin{figure}[tb]   
  \begin{center}
  \epsfig{file=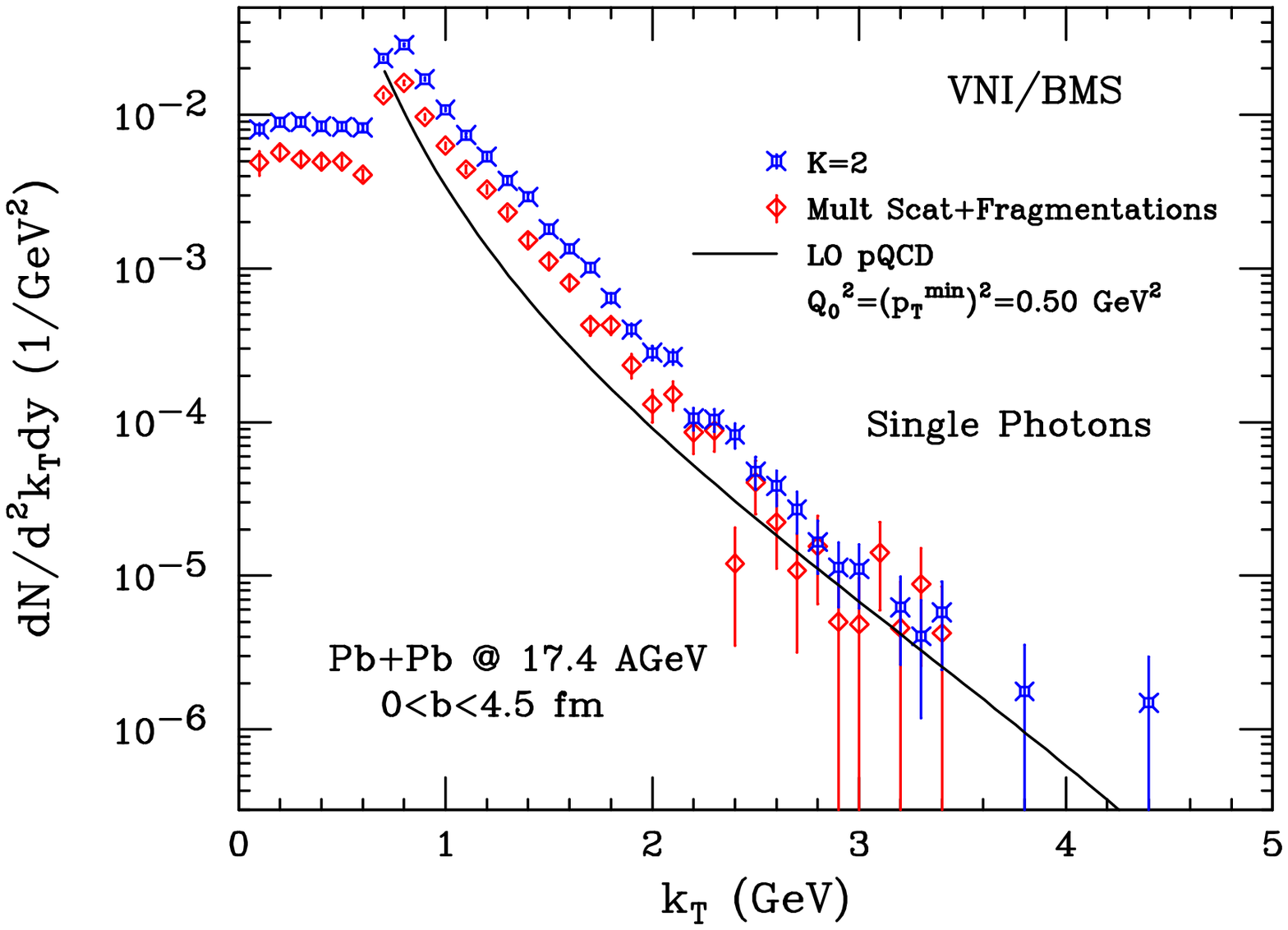,width=8.6cm}
  \epsfig{file=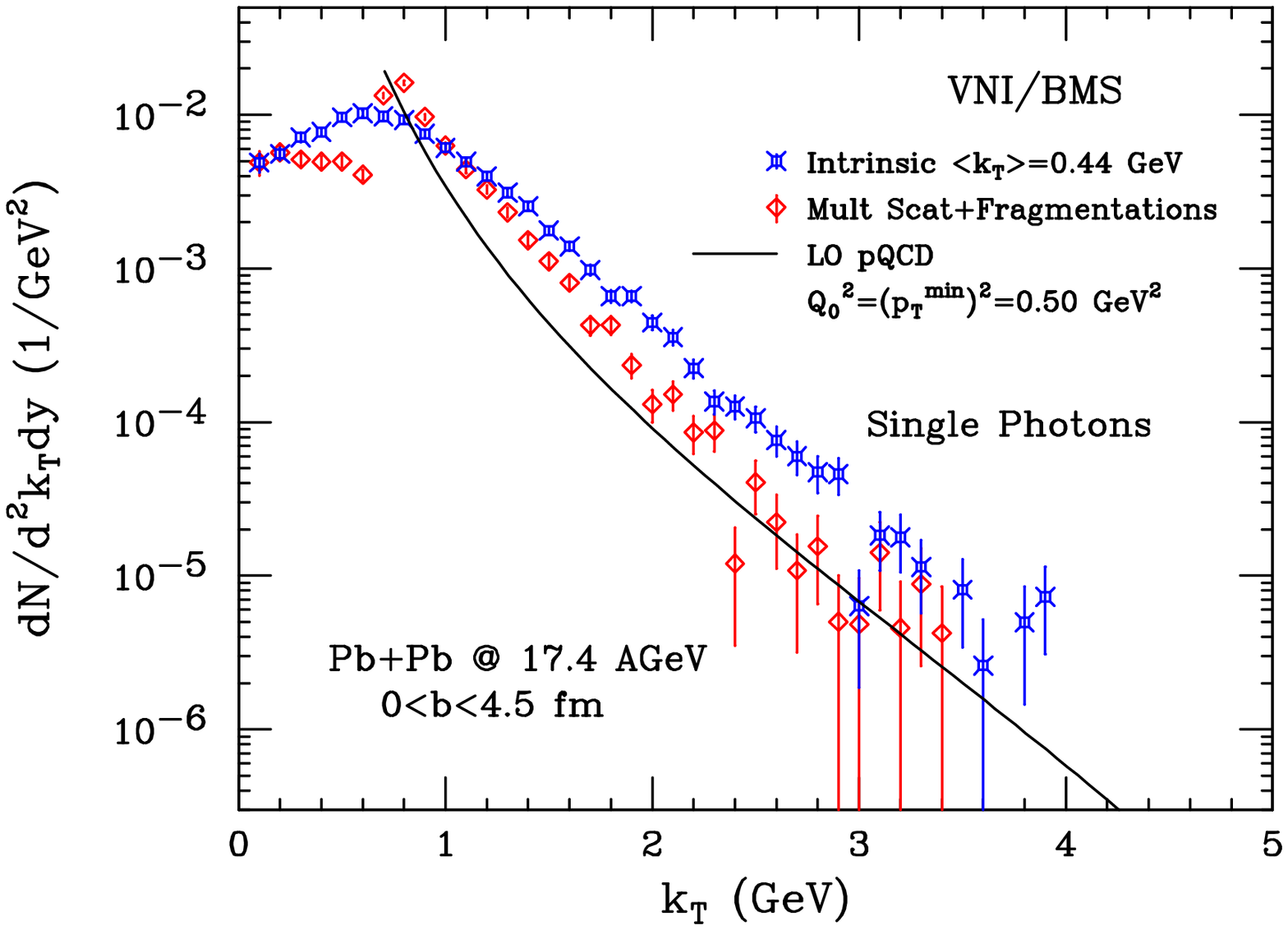,width=8.6cm}
\caption{ 
Upper panel: 
  Production of photons when multiple scatterings as well as 
  fragmentations are included, along with a $K$-factor of 2.
Lower panel: 
  Production of photons when an intrinsic $k_T$ is assigned to 
  partons. The diamonds in both cases give the results when the 
  multiple scattering and fragmentations, but no $K$-factor and 
  no intrinsic $k_T$ were included.}
\label{fig3}
\end{center}
\end{figure}

The effect of higher order corrections via an effective $K$-factor
or inclusion of an intrinsic $k_T$ in the momenta of the partons is
investigated in Fig.~\ref{fig3}. 
%
%
%
The intrinsic $k_T$ is sampled from a distribution $\sim \exp(-k_T/p_0)$,
with $p_0 =$~0.44~GeV~\cite{vni}.
While the inclusion of $K$-factor 
increases the yield of photons roughly in proportion (here $K=2$), 
the inclusion of intrinsic $k_T$ affects the results in a more complex 
manner: it reduces the yield of photons with $p_T < 1.2$ GeV, but 
increases the yield of photons with larger $p_T$ by up to a factor 
of 2 (we do not consider the results below $p_T^{\rm min}$ which are
strongly affected by the cut-off). A similar observation has been
made in \cite{dumi_phot}, where the importance of nuclear broadening
effects for the understanding of the WA98 photon data at the SPS
was emphasized.
 
\begin{figure}[tb]
\begin{center}  
\epsfig{file=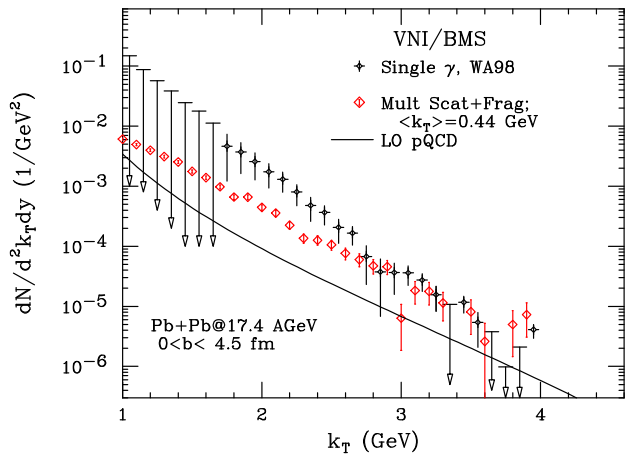,width=8.2cm}
\caption{A comparison of single photon production in the parton 
  cascade model, including multiple scatterings and fragmentations 
  of partons along with an intrinsic partonic transverse momentum, 
  characterized by $\langle k_T \rangle $ =0.44 GeV, with the WA98 
  data.}
\label{fig4}
\end{center}
\end{figure}

Figure~\ref{fig4} shows the comparison of our most complete scenario 
(parton re-scattering, fragmentation and intrinsic $k_T$) with data
from the WA98 experiment \cite{wa98}. The PCM is able to account
for the measured direct photon yield for $k_T \ge 2.7$~GeV. For
smaller $k_T$ the PCM falls short of the data, which is not surprising
since hadronic processes not included in the PCM dominate the photon 
yield in this range~\cite{photons}.

In summary, 
our calculations are able to describe the observed photon yield at 
$k_T \ge 2.7$~GeV measured by WA98 if higher-order corrections in the
form of intrinsic $k_T$ are taken into account.  Only a small increase 
in the photon yield is seen when the scattered partons are allowed 
to fragment (mostly due to $q \to q\gamma$ reactions), since the number 
of secondary collisions does {\em not} change as the fragmentations are 
rather few.  We conclude that the parton medium is dilute and does not
form a dense parton plasma, at SPS.  This is in stark contrast to the results 
obtained at RHIC energies~\cite{bms_phot} where the multiple scatterings 
and fragmentations increase the production of photons manifold over the 
prompt photon yield.  Our results suggest that the observed absence of 
jet quenching at SPS is a consequence of the modest amount of cascading 
and fragmentation of the scattered partons, while these processes occur 
in abundance at RHIC energies, resulting in a significant suppression of
high-momentum hadrons due to multiple scattering.

\begin{acknowledgments}  
This work was supported in part by RIKEN, Brookhaven
National Laboratory and DOE grants DE-FG02-96ER40945 as well as
DE-AC02-98CH10886 and the Natural Sciences and Engineering Research
Council of Canada. We thank Charles Gale for useful comments. 
\end{acknowledgments}

\end{document}